\documentclass{aa}
\usepackage{psfig}

\include{psfig}

\begin{document}


\title{An empirical method to estimate the LMC distance using B-stars
in eclipsing binary systems}

\author{ M. Salaris\inst{1,2}   \and M.A.T. Groenewegen\inst{3,4}}

\offprints{Maurizio Salaris, e-mail: ms@astro.livjm.ac.uk}

\institute {
Astrophysics Research Institute, Liverpool John Moores
University, Twelve Quays House, Egerton Wharf, Birkenhead CH41 1LD, UK
\and
Max-Planck-Institut f\"ur Astrophysik,
Karl-Schwarzschild-Stra{\ss}e 1, D-85748 Garching, Germany
\and
Instituut voor Sterrenkunde, PACS-ICC,
Celestijnenlaan 200B, B-3001 Heverlee, Belgium
\and
European Southern Observatory, EIS-team, 
Karl-Schwarzschild-Stra{\ss}e 2, D-85740 Garching, Germany
}

\date{received,  accepted}

\authorrunning{Salaris \& Groenewegen}
\titlerunning{An empirical method to estimate the LMC distance}

\abstract{We present a new method to determine the distance to
B-stars in eclipsing binary systems. The method is completely empirical,
and it is based on the existence of a very tight linear relationship 
between the $V$-band ``zero magnitude angular diameter'' and the Str\"omgren 
colour index $c_{1}$ for B-stars; we have empirically calibrated 
this relationship using local single B-stars with accurate angular 
diameters, and B-stars in eclipsing binaries with precise radii and parallax
determinations. By studying the differential behaviour of this
relationship as predicted by theoretical stellar evolution models, we
find that it is independent of the stellar metallicities for a range
of [Fe/H] values between the solar one and that of young
stars in the Magellanic Clouds.  The method, which also provides the
value of the reddening to the system, is discussed in detail, together
with a thorough estimate of the associated errors. We conclude that
accurate Str\"omgren photometry obtainable with 1.5 m-class telescopes
of the LMC eclipsing binaries HV 2274 and HV 982 will allow 
to obtain an empirical 
LMC distance with an accuracy of the order of 0.13 mag.
}

\maketitle

\section{Introduction} 

The distance to the large Magellanic Cloud is the cornerstone of the
extragalactic distance scale, since the zero point of both the
Cepheids and Type Ia supernovae distances is tied to the LMC
distance. Disappointingly enough, existing determinations of this
fundamental quantity span a wide range, comprising both ``short''
($\rm (m-M)_{0}\leq$18.30 -- e.g. Udalski 2000) and ``long'' ($\rm
(m-M)_{0}\ge$18.50 -- e.g. Feast \& Catchpole 1997) values.

In the last few years well-detached main sequence B type eclipsing
binary (EB) systems have been proposed as virtually ideal standard
candles, and employed by Guinan et al.~(1998) and Fitzpatrick et
al.~(2001 -- F01) to derive the distance to two EB systems in the LMC
(HV 2274 and HV 982). From an observational point of view, what is
required is the monitoring of the light curve (preferably in several
filters) and of the radial velocity.  Analysing this data gives the
orbital elements, the masses and radii of both components.  In
addition to that, UV/optical spectra are necessary; once obtained, a
multi-parametric fit of theoretical spectra to the observed spectrum
and the broadband data is performed, in order to simultaneously derive
the effective temperature of both components, reddening, metallicity,
micro-turbulent velocity, the distance and 5 parameters describing the
UV/optical extinction curve.  When applying this method to HV2274,
Guinan et al.~(1998) and F01 found, respectively, 
$\rm (m-M)_{0} =$ 18.30 and $\rm (m-M)_{0} =$ 18.36, with an error on the
individual determinations of the order of 0.10 mag; F01 obtained  
$\rm (m-M)_{0} =$ 18.50$\pm$0.06 for HV982. As shown
by Groenewegen \& Salaris (2001 -- GS01) for the case of HV 2274, this
method heavily relies on the absolute value of the fluxes predicted by
the theory, which, given the current uncertainty of stellar
atmospheres modelling (see GS01 for more details), can introduce large
systematic errors.  Moreover, GS01 have shown that the resulting
distance is also dependent on what broadband colours are included in
the fitting procedure, so that, even if a formal error of the order of
0.10 mag can be derived from the fitting procedure, the real
uncertainty is probably much larger.

An alternative method to use EB systems as distance indicators is
based on empirical relationships linking colours to surface brightness
(see, e.g., Barnes \& Evans 1976, Lacy 1977, Di Benedetto 1998 and
references therein). Once the colours of the individual components of
the system are known from the light curve analysis (e.g. Lacy 1977),
these relationships provide the apparent radii; since the true radii
are also known from the analysis of the radial velocity curve, the
distance to the system is derived straightforwardly. This method has
been applied recently by Thompson et al.~(2001) to one EB system in
the turn-off region of the Colour-Magnitude-Diagram of the globular
cluster $\omega$ Centauri. They made use of surface brightness (in the
$V$-band) versus colour relationships in $(V-K), (V-H), (V-J)$,
calibrated on the stellar sample by Di Benedetto~(1998), for A,F,G,K
dwarfs and giants.  This method is very much appealing, since it
avoids the use of the still uncertain model atmospheres. One drawback
is that it is necessary to know in advance the reddening of the
system, which is not an easy task if one wants to apply this technique
to LMC objects.  Moreover a calibration for B-stars -- like the stars
in the LMC EB systems previously discussed -- does not exist yet.

In this paper we propose a variant of this surface brightness
technique using Str\"omgren colours, suited to determine the distance
of B-stars in EB systems. This method turns out to be largely
independent of the metallicity of the components and, moreover,
working with the Str\"omgren filters one can easily derive the
reddening of B-stars from the $c_{1}-(b-y)$ diagram (see, e.g.,
Larsen, Clausen \& Storm~2000 -- LCS00).

In Sect.~2 we discuss the theoretical background of the method, while
in Sect.~3 we present our calibration and use for B-stars.  Section~4
discusses the errors associated with the method through an application
to some Galactic EB systems and the LMC system HV 982.  Conclusions
follow in Sect.~5.

\section{Theoretical background}

The surface brightness ($S$) of a star is classically defined as:

\begin{equation}
S_{\lambda} = m_{\lambda} + 5 \log \phi
\end{equation}
where $\phi$ is the angular diameter in milli-arcseconds, and
$m_{\lambda}$ represents the apparent magnitude in a given passband.
$S_{\lambda}$ is determined empirically by measuring the stellar
angular diameters and brightness of non-variable stars.

A slightly alternative concept, but essentially equivalent, is used by
van Belle (1999) to define the angular diameter a star would have if
its magnitude (in a given passband) were zero (zero magnitude angular
diameter):
\begin{equation}
{\phi}_{(m_{\lambda}=0)} = \phi \; 10^{ m_{\lambda} / 5}
\end{equation}
where $\phi$ is again the angular diameter.

In either case, $S_{\lambda}$ and ${\phi}_{(m_{\lambda}=0)}$ are
empirically calibrated against a colour ($m_{2} - m_{3}$) on a sample
of local stars.  In the case of well-detached EBs, magnitudes and
colours for the individual components are derived using information
from the light-curve analysis, and dereddened.  The observed colours
and magnitudes, together with the empirical relationships previously
discussed, provide the angular diameters (in mas) of the two
components.  From the angular diameter and the absolute radius ($R$)
obtained from the analysis of the radial velocity curve, one obtains
easily the distance to the two components according to the
relationship:

\begin{equation}
d({\rm pc}) = 1.337 \times 10^{-5} \; R({\rm km})/\phi({\rm mas})
\end{equation}

It is obvious that in order to apply this method one needs to know the
radius of the stars, therefore it can be reliably applied only to EB
systems (but see also Allende Prieto~2001). In principle one can use
any magnitude and colour in conjunction with Eqs.~1 or 2.

\section{Calibration for B-stars with Str\"omgren photometry}

When trying to apply surface brightness techniques one must face the
problem of reddening. The reddening in, e.g., the LMC is strongly
variable on very short spatial scales (see, e.g., Romaniello et
al.~2000, LCS00), and the ideal way to proceed is to determine it on a
star-by-star basis.  Str\"omgren $uvby$ photometry provides the tool
to do this, at least for B-stars, by employing the $c_{1}-(b-y)$
diagram (We briefly recall that $c_{1}=(u-v)-(v-b)$).  As discussed by
Larsen et al.~(2000), in this plane luminosity class IV and V B-stars
are located on a standard sequence (see Table~1) which is largely
independent of the star metallicity (LCS00).  In Fig.~1 we show the
standard sequence by Perry, Olsen \& Crawford~(1987), and the
direction of the reddening vector, which is nearly horizontal, since
$(c_{1})_0=c_{1}-0.20E(b-y)$. To first order, the reddening $E(b-y)$
can be estimated as simply being the difference between the observed
$(b-y)$ and the intrinsic value $(b-y)_{0}$ corresponding to the
observed $c_{1}$. We recall that $E(B-V)=1.4E(b-y)$.

\begin{figure}
\psfig{figure=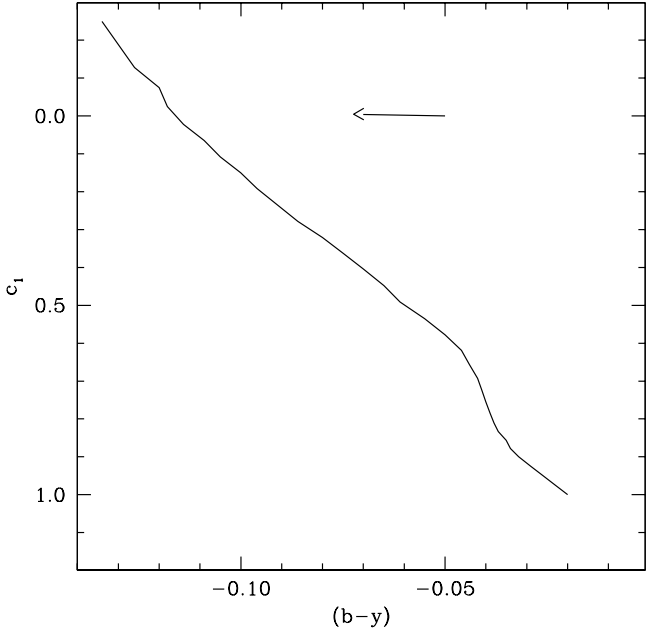,width=9.0cm,clip=}
\caption[]{Standard sequence for B-stars. The arrow shows the 
direction of the reddening vector.}
\protect\label{c0by}
\end{figure}

To confirm the independence of this standard sequence on metallicity,
we show in Fig.~2 the $c_{1}-(b-y)$ relationship for B-stars between
the zero age main sequence and the turn off (luminosity classes IV and
V) with ages between 10 and 50 Myr (the precise age range is not
relevant), from the theoretical models by Bertelli et al.~(1994),
transformed to the Str\"omgren filters by means of the Kurucz (1992)
colour transformations. In the [Fe/H] range displayed -- [Fe/H]
between 0.0 and $-$0.7, which covers the metallicity range of B-stars
in the Milky Way, LMC and SMC (see, e.g., the discussion in GC01) --
the standard sequence of B-stars is practically unaffected by the
stellar chemical composition. The maximum difference is less
than 0.01 mag in $(b-y)$ at a given value of $c_{1}$ between the
[Fe/H]=$-0.7$ and 0.0 sequence.


Therefore, as a first step, after the radial velocity and light curve
analysis (from Str\"omgren filter observations) of a given EB system
with B main sequence star components, one can derive the reddening of
the two objects by simply using the $c_{1}-(b-y)$ diagram and the
sequence provided in Table~1.

\begin{table}
\caption{Standard sequence for B-stars 
(from Perry et al.~1987).} 
\protect\label{t:redd}
\begin{tabular}{cccc}
\hline
$(b-y)$ & $c_{1}$ &$(b-y)$ & $c_{1}$\\
\hline
$-$0.134 &$-$0.250 &$-$0.050 &  0.578\\
$-$0.126 &$-$0.128 &$-$0.046 &  0.619\\
$-$0.120 &$-$0.075 &$-$0.044 &  0.656\\
$-$0.118 &$-$0.025 &$-$0.042 &  0.693\\
$-$0.114 &  0.022 & $-$0.041 &  0.724\\
$-$0.109 &  0.065 & $-$0.040 &  0.755\\
$-$0.105 &  0.108 & $-$0.039 &  0.785\\
$-$0.100 &  0.150 & $-$0.038 &  0.811\\
$-$0.096 &  0.192 & $-$0.037 &  0.833\\
$-$0.091 &  0.235 & $-$0.035 &  0.856\\
$-$0.086 &  0.278 & $-$0.034 &  0.878\\
$-$0.080 &  0.321 & $-$0.032 &  0.900\\
$-$0.075 &  0.362 & $-$0.029 &  0.925\\
$-$0.070 &  0.404 & $-$0.026 &  0.950\\
$-$0.065 &  0.448 & $-$0.023 &  0.975\\
$-$0.061 &  0.491 & $-$0.020 &  1.000\\
$-$0.055 &  0.535 &        &       \\
\hline
\end{tabular}
\end{table}

\begin{table*}[t]
\caption{Calibrating stars for the ${\phi}_{(V=0)}$-$c_{1}$
relationship. The photometric data ($c_{1}$, $(b-y)$
and apparent magnitudes $V$) have been de-reddened 
(if needed) using the standard sequence in Table~1.} 
\protect\label{t:calib}
\begin{tabular}{cccccl}
\hline
${\phi}_{(V=0)}$ & $c_{1}$ &$(b-y)$ & $V$ & E($b-y$) & name\\
\hline
1.53 $\pm$ 0.09  &0.111$\pm$0.020 & $-$0.091$\pm$0.020 & 1.64$\pm$0.02
& 0.00 & HD  35468\\
2.24 $\pm$ 0.20 &0.490$\pm$0.030 & $-$0.070$\pm$0.020 & 4.30$\pm$0.01 &
0.00 &  HD  6882A\\ 
2.74 $\pm$ 0.24 &0.770$\pm$0.030 & $-$0.010$\pm$0.020& 5.68$\pm$0.01 &
0.00  &  HD  6882B\\
2.55 $\pm$ 0.12  &0.712$\pm$0.020 & $-$0.041$\pm$0.020 & 1.35$\pm$0.02
&0.00 & HD  87901\\
2.05 $\pm$ 0.28  &0.447$\pm$0.012 & $-$0.065$\pm$0.007 &
5.83$\pm$0.009& 0.01 & HD 135876A\\   
2.61 $\pm$ 0.35  &0.807$\pm$0.030 & $-$0.038$\pm$0.019 &
7.07$\pm$0.014 & 0.01 & HD 135876B\\ 
1.10 $\pm$ 0.12 &$-0.037\pm$0.020 & $-$0.119$\pm$0.020 & 1.88$\pm$0.02
& 0.10 & HD 143275\\
1.96 $\pm$ 0.13  &0.271$\pm$0.020& $-$0.092$\pm$0.020 & 1.94$\pm$0.02
& 0.00 & HD 193924\\
2.27 $\pm$ 0.16  &0.567$\pm$0.020& $-$0.058$\pm$0.020 & 1.74$\pm$0.02
& 0.00 &  HD 209952\\
\hline
\end{tabular}
\end{table*}

\begin{figure}
\psfig{figure=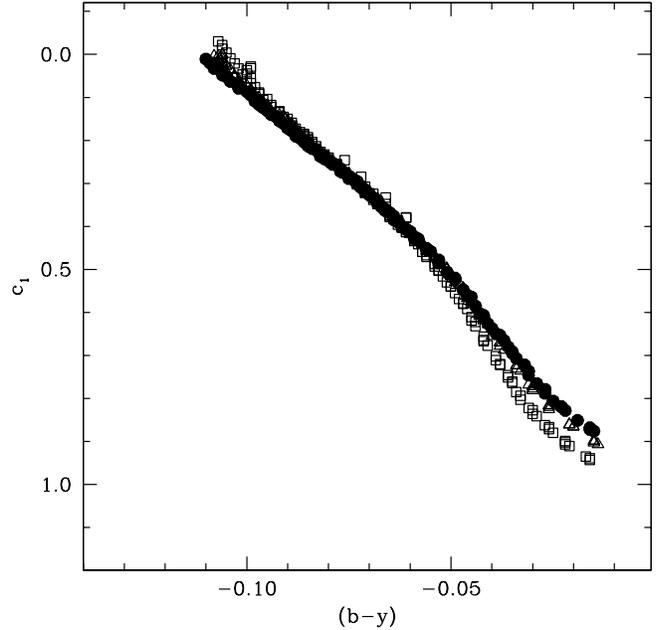,width=9.0cm,clip=}
\caption[]{Theoretical sequences for main sequence B-stars
with ages between 10 and 50 Myr and
[Fe/H]=$-$0.7 (filled circles), $-$0.4 (open triangles) 
and 0.0 (open squares).}
\protect\label{c0bytheo}
\end{figure}

As a second step, one needs a relationship between the surface
brightness or zero magnitude angular diameter, and a de-reddened
colour. The following precepts have guided our choice:
\begin{itemize}

\item the relationship should be tight and mathematically simple

\item there has to be a sufficient number of local calibrators

\item it has to be dependent as weakly as possible on photometric
errors, reddening uncertainties, and differences in the chemical
composition between calibrating and observed objects 

\end{itemize}

We found that a relationship fulfilling all these constraints does
exist for B-stars, and links ${\phi}_{(V=0)}$ with $c_{1}$.  In
Table~2 we provide the values of ${\phi}_{(V=0)}$, $c_{1}$ and $(b-y)$
(reddening corrected) for the 9 calibrating B-stars we found in the
literature; they cover fairly well the entire range of B main sequence
stars.  Four of these objects are in 2 EB systems with well determined
$Hipparcos$ parallaxes -- ${\sigma}_{\pi}/\pi<12\%$ -- (photometry and
radii are taken from Jordi et al.~1997, and Ribas et al.~1998) and
radii with an error of less than 2\%; their absolute radii have been
therefore transformed into angular diameters as if they were located
at a distance of 10 pc, which have been then used, together with the
absolute $V$ magnitudes, to compute ${\phi}_{(V=0)}$\footnote{Data for
other EB systems were available with the appropriate photometric and
radius errors, but the uncertainty on their parallaxes was higher,
thus causing a very large uncertainty on their ${\phi}_{(V=0)}$
values. Since the 2 systems we choose have been selected on the base
of their parallax error, we applied the Lutz-Kelker correction to the
distances obtained from their parallax, following Smith (1987), and
assuming a uniform stellar distribution. The corrections applied
to the absolute V magnitudes of the EB systems are $-$0.046 and
$-$0.149 mag for, respectively, HD 6882 and HD 135876.}.  
The remaining stars have direct angular diameter
determinations accurate to better than 9\% from Code et al.~(1976),
and photometry obtained from the Mermilliod, Mermilliod \&
Hauck~(1997) catalogue.  Errors on the photometric data come from
Jordi et al.~(1997) for the EB systems, while we have adopted errors
of 0.02 mag for the data obtained from the Mermilliod, Mermilliod \&
Hauck~(1997) catalogue.

In Fig.~3 we show the calibration of the empirical 
$\phi_{(V=0)}-c_{1}$ relationship using the data in Tab.~2, 
which is given by
\begin{equation}
$$\phi_{(V=0)} = 1.824 (\pm 0.180)\; c_{1} + 1.294 (\pm 0.078)$$
\end{equation}
with a linear correlation coefficient r=0.99 (corresponding to a
probability of $\sim$1.0e-6 that the points lie on this relationship
only by chance).

\begin{figure}
\psfig{figure=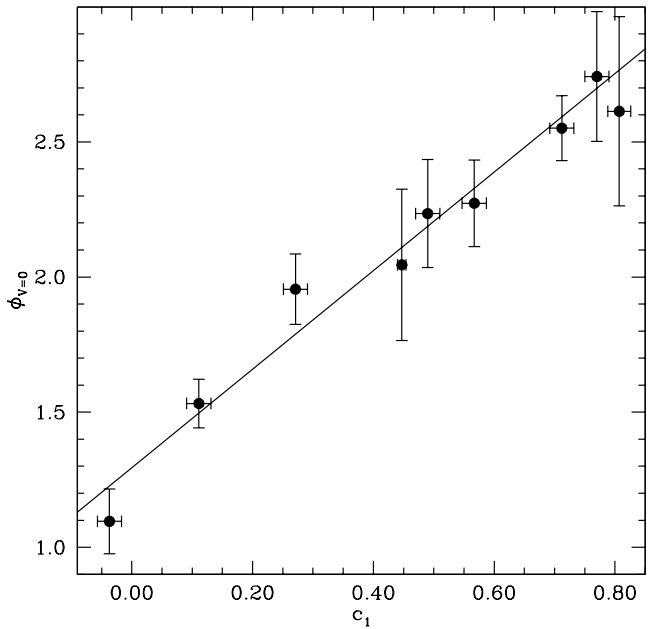,width=9.0cm,clip=}
\caption[]{Calibration of the ${\phi}_{(V=0)}-c_{1}$
relationship. Points are the empirical data, while the solid line shows
the best fit to the data (Eq.~4).}
\protect\label{calib}
\end{figure}

\begin{table*}[t]
\caption{Data for the Galactic systems analyzed (see text for details).} 
\protect\label{t:examples}
\begin{tabular}{lcrccccc}
\hline
name & $V$ & $(b-y)$ & $c_{1}$&
radius & $E(b-y)$ & $(m-M)_{0}$ & $(m-M)_{0. Hip}$ \\
\hline
HD24909A & 7.866$\pm$0.014 & 0.056$\pm$0.004& 0.635$\pm$0.011&
2.446$\pm$0.026 $\rm R_{\odot}$&
0.102$\pm$0.010 & 7.29$\pm$0.12 & 7.62$\pm$0.78\\
HD161783A & 6.177$\pm$0.010 &$-$0.038$\pm$0.005& 0.249$\pm$0.010&
4.432$\pm$0.084 $\rm R_{\odot}$&
0.053$\pm$0.010 & 7.83$\pm$0.12 & 7.01$\pm$0.52\\
HD218066A & 8.35$\pm$0.04 & 0.333$\pm$0.010& 0.037$\pm$0.015&
5.685$\pm$0.130 $\rm R_{\odot}$&
0.455$\pm$0.010 & 9.60$\pm$0.16 & 9.30$\pm$1.53\\
HD218066B & 8.60$\pm$0.04 &0.339$\pm$0.010& 0.045$\pm$0.015&
5.177$\pm$0.129 $\rm R_{\odot}$&
0.455$\pm$0.010 & 9.62$\pm$0.16 & 9.30$\pm$1.53\\
\hline
\end{tabular}
\end{table*}

At present there are no calibrations of the
relationship between zero magnitude angular diameter (or surface brightness)
and colours, spanning the entire main sequence B star range, other than 
the one we obtained. Since there are essentially 
no infrared colours available for our calibrating objects, 
neither the Di Benedetto~(1998), nor the van
Belle~(1999) calibrations cover this range. 
It would be very interesting, when data will be
available in the future, to compare our result with equivalent
calibrations using, e.g., (V-K).

An important issue is the dependence of Eq.~4 on the
stellar metal content, since the calibrators are local stars of
presumably solar-like metallicity, while the B-stars in the LMC are
more metal poor. The only way to assess this dependence, due to the
lack of appropriate data for more metal poor stars, is -- as in the
case of the $c_{1}-(b-y)$ standard sequence -- to use the differential
(not the absolute quantities) behaviour of theoretical models.
Figure~4 shows in the $\phi_{(V=0)}-c_{1}$ plane the same models
displayed in Fig.~2.

For B-stars of luminosity class IV and V one expects that the
$\phi_{(V=0)}-c_{1}$ relationship is negligibly affected by the
metallicity for [Fe/H] between 0.0 and $-$0.7. In the case of
[Fe/H]=$-0.7$ (typical metallicity of SMC young stars) the difference
with respect to the solar case is small for late type B-stars (higher
values of $c_{1}$), causing  a systematic underestimate of the
distance (when using the local calibration) of the order of only 0.05 mag, 
and completely vanishing for earlier types.  It is therefore possible to
apply Eq.~4 with confidence also to stars in the Magellanic Clouds.

Summarizing, from the Johnson $V$ and Str\"omgren $uvby$ light curves
of B-stars in well-detached EB systems, one can derive the individual
$V$ magnitude and Str\"omgren colour of the single components.  Using
the standard relationship in Table~1, the reddening of the system can
be easily estimated. Once the de-reddened $c_{1}$ values of the 2
components are at hand, one can employ Eq.~4 to derive the
corresponding zero magnitude angular diameters in V, which can be
transformed to true angular diameters using Eq.~2.  At this stage,
since the angular diameters and the physical radii (derived from the
analysis of the radial velocity curve) are known, one can finally use
Eq.~3 to obtain the distance of the individual stars.  As a check for
the consistency of the results, one has to ensure that the distances
of the two components are in agreement within the errors, or,
equivalently, that the ratio of the derived angular diameters agrees
with the observed radii ratio.

\begin{table*}[t]
\caption{Data for HV 982} 
\protect\label{t:HV982}
\begin{tabular}{lccccccc}
\hline
name & $V$ & $(b-y)$ & $c_{1}$&
radius & $E(b-y)$ & $(m-M)_{0}$\\
\hline
HV 982 A & 15.26$\pm$0.02 & 0.008$\pm$0.041& 0.062$\pm$0.045&
8.02$\pm$0.23 $\rm R_{\odot}$&
0.120$\pm$0.045 & 18.43$\pm$0.24\\
HV 982 B & 15.48$\pm$0.02 & 0.010$\pm$0.041& 0.076$\pm$0.045&
7.35$\pm$0.21 $\rm R_{\odot}$&
0.120$\pm$0.045 & 18.42$\pm$0.24\\
\hline
\end{tabular}
\end{table*}

\begin{figure}
\psfig{figure=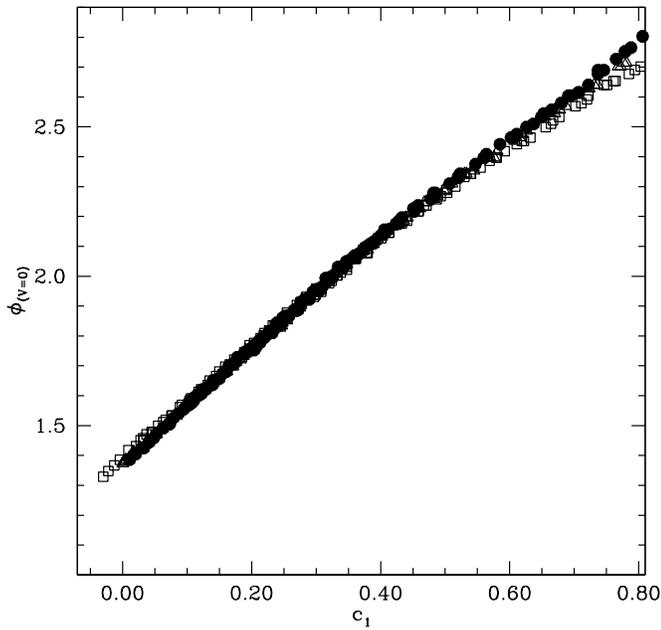,width=9.0cm,clip=}
\caption[]{Theoretical ${\phi}_{(V=0)}-c_{1}$
relationships for [Fe/H]=$-$0.7 (filled circles), 
$-$0.4 (open triangles) and 0.0 (open squares).}
\protect\label{theory}
\end{figure}

\section{Error estimate and constraints on LMC observations}

In this section we will apply the previously outlined method to some
B-stars in Galactic EB systems, in order to estimate the associated
errors.  We have considered 3 EB systems spanning the entire spectral
range of B main sequence stars; the data about these systems are
displayed in Tab.~3. Observed magnitudes and colours (columns 2, 3 and
4) come from the compilation by Jordi et al.~(1997), with the
exception of the error bars on the Str\"omgren data for HD161783,
which we have computed from the original paper by Clausen (1996);
radii (column 5) come from the compilation by Andersen (1991), while
columns 6 and 7 show reddenings and distance moduli obtained with our
technique. Finally, column 8 displays the distance according to the
Hipparcos parallax.  For HD24909 and HD161783 we considered just one
component, since for the former system the secondary star is an
A-star, while for the latter one the secondary star is variable.  All
3 systems have parallaxes with large error bars; otherwise, we would
have included them in our empirical calibration of Eq.~4.  We want
also to mention the fact that the $c_{1}$ values for both components
of HD218066, once de-reddened, are slightly outside the range of
validity of our calibration (by less than 0.02 mag).  The stars in
Tab.~3 are ordered according to their spectral type; HD24909A is a
late type B-star, while the components of the HD218066 system are both
early type B-stars. Notice the good agreement between the distances
derived for the components of this system.

The errors on the distances have been derived by adding in quadrature
the contributions of the errors on $V$, $(b-y)$, $c_{1}$, the
individual radii and the coefficients of Eq.~4.  The largest component
of the total error comes from the errors on the coefficients of Eq.~4;
they alone provide an uncertainty of about 0.10-0.12 mag on the final
error on the distance modulus.  Errors on the photometry and on the
radii are of much less importance, as long as they are within 
$\sim$ 0.02 mag and 3\%, respectively.

To summarize the results of these error estimates, our empirical
method to derive distances to individual B-stars in EB systems
provides distance moduli with errors of the order of 0.12 -- 0.16 mag,
when applied to systems with photometry and radius determinations
precise at the level discussed before.

In case of the LMC, HV 2274 and HV 982 are EB systems with early type
B-stars, for which radius determinations and $V$ photometry have a
high enough precision (errors of the order of 3\% on the radii and
0.02 mag on V) to make the error on the distance dominated only by the
dispersion associated to Eq.~4.  Unfortunately, Str\"omgren data for
HV 2274 are not available in the literature, while for HV982 the
Str\"omgren photometry by Pritchard et al.~(1998 -- P98) has large
errors of the order of 0.04--0.05 mag in $(b-y)$ and $c_{1}$.
Nevertheless, as an exercise, we derive the distance to HV 982
following our method. In Tab.~4 we present the observational data for
this system (Str\"omgren data are from P98, $V$-band data and radii
are from F01) and the derived reddening and distance modulus of both
components.

The derived reddening is in agreement with the estimate for B-stars in
a field around HV 982 derived by LCS00 using the $(b-y)$ and $c_{1}$
indices; they find an average E$(b-y)$ = 0.10 with values ranging
between $\sim$0.04 and $\sim$0.25 mag. FP01 derive a reddening of
E$(B-V)$ = 0.086 $\pm$ 0.005 from their multidimensional spectral
fitting method, consistent with the value derived here. 
The distances
we obtain, $(m-M)_{0}$=18.43$\pm$0.24 and $(m-M)_{0}$=18.42$\pm$0.24,
do not put very strong constraints on the LMC distance, given the
large error bar.
However, it is feasible to get Str\"omgren photometry accurate to
0.02-0.03 mag in $(b-y)$ and $c_{1}$ with only 1.5 m-class telescopes
(see, e.g., the data by LCS00); already with these errors one could
get the distance to HV 982, for example, with an accuracy of 0.17 --
0.18 mag, equivalent to a parallax error of about 8\%. Since both
HV 982 and HV 2274 lie near the LMC center (van der Marel \& Cioni
2001), if Str\"omgren data of the same precision were available also
for HV 2274, it would be possible to combine both distances and obtain
an empirical LMC distance accurate to 0.13 mag.

\section{Conclusions}

We have presented a new technique to estimate reddening and distance
moduli to B-stars in EB systems.  It is based on the observation of
the $Vuvby$ light curve and the consequent determination of $V$,
$c_{1}$ and $(b-y)$ magnitudes for the individual components of the
binary. A comparison of the stars' positions on the $c_{1}-(b-y)$
plane with respect to the local sequence of Table~1 permits to derive
the reddening to the system.  After de-reddening the individual $V$
and $c_{1}$ values, the use of an empirically calibrated
${\phi}_{(V=0)}-c_{1}$ relationship (Eq.~4) allows one to derive the
apparent diameter of the objects; this apparent diameter, coupled with
the knowledge of the physical radius from the radial velocity
analysis, provides straightforwardly the distance to the system
(Eq.~3).  These two empirical relationships used to derive reddening
and distance are expected to be basically independent of the stellar
metallicities in the [Fe/H] range covered by young stars in the solar
neighborood, LMC and SMC.  Photometric errors within $\sim$0.02 mag
and radius uncertainties up to 3\% allow to determine distances to
individual systems with an error bar of 0.12 -- 0.17 mag. If this
photometric accuracy is obtained for the photometry of the two LMC
systems HV 2274 and HV 982 -- which have both precise radius
determinations and lie close to the LMC center -- one can derive an
empirical estimate of the LMC distance with an accuracy of the order
of 0.13 mag.

\acknowledgements{
M.S. wishes to thank the Max-Planck-Institut f\"ur Astrophysik for 
the kind hospitality during the completion of this work. We 
thank Phil James for comments on a preliminary version of the manuscript. 
This research has made use of the {\sc simbad} database, operated at
CDS, Strasbourg, France.}

{}

\end{document}